\documentclass{llncs}

\usepackage[latin9]{inputenc}
\usepackage[T1]{fontenc}
\usepackage{multicol}
\usepackage{url}
\usepackage{graphicx}
\usepackage[FIGBOTCAP]{subfigure}
\usepackage{multirow}
\usepackage{boxedminipage}

\def\|#1|{\texttt{#1}}

\begin{document}

\title{Building Portable Thread Schedulers for Hierarchical
Multiprocessors: the BubbleSched Framework}

\author{Samuel Thibault \and
Raymond Namyst \and
Pierre-André Wacrenier
}

\institute{
INRIA Futurs - \textsc{LaBRI} ---
351 cours de la libération ---
33405 Talence cedex, France\\
\email{\{thibault,namyst,wacrenier\}@labri.fr}
}

\maketitle

\begin{abstract}
  Exploiting full computational power of current more and more
  hierarchical multiprocessor machines requires a very careful
  distribution of threads and data among the underlying non-uniform
  architecture. Unfortunately, most operating systems only provide a
  poor scheduling API that does not allow applications to transmit
  valuable \emph{scheduling hints} to the system. In a previous
  paper~\cite{THIBAULT:2005:31780}, we showed that using a
  \emph{bubble}-based thread scheduler can significantly improve
  applications' performance in a portable way. However, since
  multithreaded applications have various scheduling requirements,
  there is no universal scheduler that could meet all these needs. In
  this paper, we present a framework that allows scheduling experts to
  implement and experiment with customized thread schedulers. It
  provides a powerful API for dynamically distributing bubbles among
  the machine in a high-level, portable, and efficient way. Several
  examples show how experts can then develop, debug and tune their own
  portable \emph{bubble schedulers}.

\end{abstract}

\section{Introduction}

Both \textsc{AMD} and \textsc{Intel} now provide quad-core chips and
are heading for 100-cores chips. This is, in the
low-end market, the emerging part of a deep trend, in the scientific
computation market, towards more and more complex machines
(e.g.~\textsc{Sun WildFire}, \textsc{SGI Altix}, \textsc{Bull
  NovaScale}).  Such large shared-memory machines are typically based
on Non-Uniform Memory Architectures (NUMA). Recent technologies such
as Simultaneous Multi-Threading (SMT) and multi-core chips make these
architectures even more hierarchical.

Exploiting these machines efficiently is a real challenge, and a
thread scheduler is faced with dilemmas when trying to take into
account the memory hierarchy and the CPU utilization
simultaneously. On NUMA machines for instance, threads should
generally be scheduled as close to their data as possible, but
bandwidth-consuming threads should rather be distributed over
different chips.  The core scheduler of operating systems can often be
influenced, but it misses the precise application behavior: for
instance, adaptively-refined meshes entail very irregular and
unpredictable behavior.  A good solution would be to let application
programmers take control of the scheduling, but writing a whole
scheduler for hierarchical machines is a very difficult task.

In a previous paper~\cite{THIBAULT:2005:31780}, we introduced the
\emph{bubble} scheduling concept that helps to express the inherent
parallel structure of multithreaded applications
in a way that can be efficiently exploited by the
underlying thread scheduler. \emph{Bubbles} are
abstractions to group threads which ``work together'' in a
recursive way.  The first \emph{proof-of-concept} implementation of
our bubble scheduler was featuring a generic hard-coded scheduler.
However, applications may have different scheduling requirements and
thus may attach different semantics to \emph{bubbles}, enforcing
memory affinity or emphasizing a high frequency of global
synchronization operations for instance. Obviously, no generic
scheduler can meet all these needs.
In this paper, we present \emph{BubbleSched}, a framework designed to
ease the development and the evaluation of customized, high-level
thread schedulers.

\section{On the Design of Thread Schedulers}
\label{sec:sched-design}

Designing a thread scheduler for hierarchical machines is complex
because it means finding an application-specific compromise between lots
of constraints: favoring affinities between threads and memory, taking
advantage of all computational power, reducing synchronization cost,
etc.

\subsection{What Input Can a Scheduler Expect?}

To make appropriate decisions at execution time, a thread scheduler
can combine a number of parameters to evaluate the goodness of
each potential scheduling action. These parameters can be collected
from several places, at different times.

\emph{At runtime}, some useful knowledge about the target machine can
be discovered. The scheduler can not only get the number of processors
but also the architecture hierarchy: how processors and memory banks
are connected, how cache levels are shared between processors, etc.
Moreover, indication about how well the threads are using the
underlying processors can be fetched from performance counters. Some
NUMA chips can also report the ratio of remote memory accesses.  The
scheduler can hence check whether threads and data are properly
localized, or enable automatic migration
policies~\cite{DynPageMigUser}.

\emph{The compiler} can also provide information about the application
behavior. Data access patterns may be
analyzed~\cite{CompilerRefAffinity} and used to choose runtime
allocation policies. An OpenMP compiler can sometimes accurately
estimate the amount of data shared by the threads involved in the same
parallel section.

Last but not least, \emph{programmers} can also provide relevant
information about the application behavior: how threads will mostly
access data, what threads are I/O-bounded, etc.) Such information can
help the scheduler to find a good trade-off regarding the co-location
of threads and data, and more generally can help to determine the
combination of scheduling policies that will perform best.

To sum it up, a lot of information is at the scheduler's disposal or
can be collected at run time. Programmers can sometimes even provide
additional scheduling hints.  All we need is a reasonable way to pass
this information from the programmer to the scheduler.

\subsection{On the Importance of Scheduling Guidance}
\label{usual_sched}

\emph{Opportunistic} policies based on
Self-Scheduling~\cite{DiversSS} are the most natural approaches
regarding thread scheduling. They use a centralized
list of ready threads (\textsc{FreeBSD~4}, \textsc{Linux~2.4},
\textsc{Windows~2K}) or a distributed one
(\textsc{FreeBSD~5}, \textsc{Irix}, \textsc{Linux~2.6}) associated to
load balancing mechanisms.
Such an approach scales and adapts to new workloads, but
it does not use affinity information from the compiler or the
programmer, and thus can not achieve best performance.

For very regular problems, a \emph{predetermined approach} can be
used.  The idea is to compute \emph{a priori} a good distribution of
threads and data that will be enforced during execution. The machine
being dedicated to the application, thread scheduling can be fully
controlled by binding exactly one kernel thread to each
processor. This approach, used in the \textsc{PaStiX}
solver for sparse linear systems~\cite{PaStiX}, gives excellent
performance for regular problems, but as soon as the solving time
depends on the data or intermediate results,
behavior prediction fails and performance may actually get worse than
by using an opportunistic approach.

An intermediate approach 
is based on \emph{negotiation}. Some language extensions
such as \textsc{OpenMP}, High Performance Fortran
(HPF) or Unified Parallel C (UPC) let
programmers write parallel applications by simply annotating the
source code. The distribution and scheduling decisions then belong to
the compiler. Such extensions get good performance by exploiting
information from programmers, but the expressiveness is limited to
``\texttt{Fork-Join}'' decomposition schemes, and programmers can not
express unbalanced parallelism for instance.

Negotiation yet appears to be a promising approach, because
programmers just have to give suggestions and clues, sometimes
indirectly, about the behavior of the threads. However, the runtime
system has only little control over the operating system's thread
scheduler. We believe that a good approach is to extend the scheduler
interface so as to allow applications to transmit scheduling
hints that will persist inside the scheduler during the lifetime of
the threads.

\subsection{Towards a Toolbox for Developing Thread Schedulers}
\label{sec:toolbox}

\begin{figure}[tb]
\centering
\includegraphics[scale=0.7]{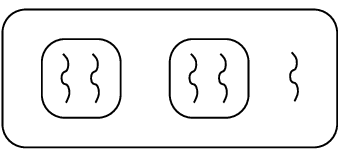}
\hspace{2cm}
\includegraphics[scale=0.7]{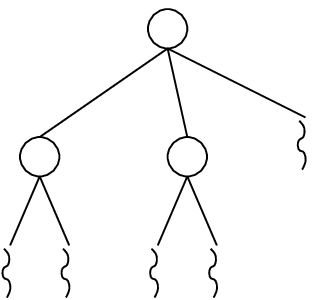}
\caption{Expressing thread relationships: graphical and tree-based
representations}
\label{ex_bubbles}
\end{figure}

In \cite{THIBAULT:2005:31780}, we proposed a model that allows
programmers to model the relationships between threads using
nested sets called \textbf{bubbles}. Figure \ref{ex_bubbles}
illustrates this: four threads are grouped as pairs in bubbles
(\emph{e.g.} they work on the same data), which are themselves grouped
along another thread in a larger bubble.
This lets express relations like data
sharing, collective operations, or more generally a particular
scheduling policy need (serialization, gang scheduling, etc.) We also
automatically model hierarchical machines with a hierarchy of
runqueues.  To each component of each hierarchical level of the machine
is associated one runqueue: one per logical processor,
core, chip, NUMA node, and one for the whole machine.
Our ground scheduler then uses a hierarchical Self-Scheduling
algorithm. Idle processors scan all runqueues that span
them, and executes the first thread that they find, from bottom to
top. For instance, if the thread is on a runqueue that represents a
chip, it may be run by any processor of this
chip.

\section{BubbleSched: A Framework for Building Portable Schedulers}
\label{sec:framework}

To tackle the delicate issue of scheduling an application on
hierarchical machines in an efficient, flexible and portable way, we
propose a new platform for easily writing customized schedulers,
based on our high-level \emph{bubble} abstractions.

Our platform allows programmers of specialized scientific libraries or
parallel programming environments to easily write thread schedulers
for various application classes on these
machines. It provides a high-level API for writing powerful and
portable schedulers that manipulate threads grouped into bubbles, as
well as tracing tools to help analyzing the dynamic behavior of these
schedulers. Programmers can hence focus on algorithmic issues rather
than on technical details.

\subsection{An API for Writing Scheduling Strategies}

\begin{figure}[tb]
\centering
\subfigure[Good CPU usage, poor affinity
care.]{
\includegraphics[scale=0.7]{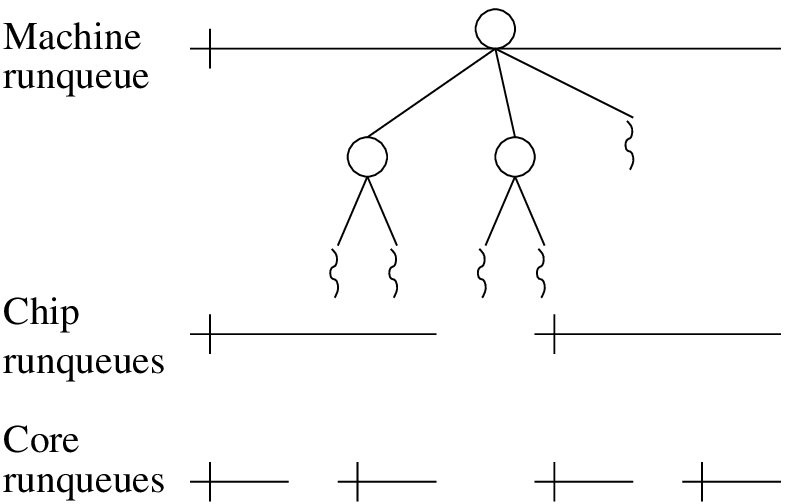}
\hspace{0.2cm}
\label{bulles_machine}
}
\subfigure[Distributed CPU usage, good affinity
care.]{
\hspace{0.2cm}
\includegraphics[scale=0.7]{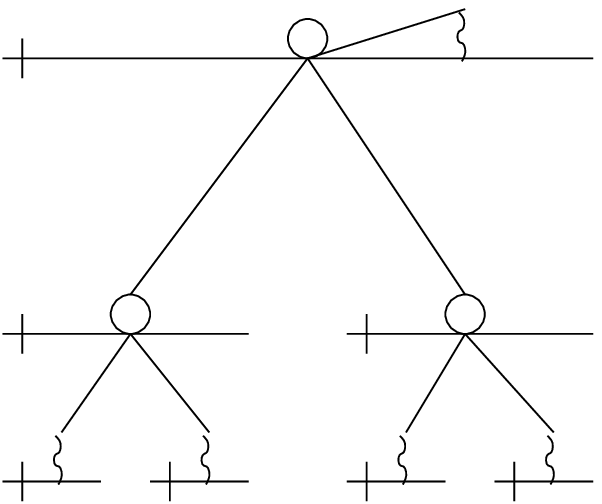}
\hspace{1cm}
\label{bulles_runqueues}
}
\caption{Possible distributions of threads and bubbles among the machine.}
\label{bulles_repart}
\end{figure}

By default, bubbles are placed
on the machine runqueue upon start-up.  In the previous example of
Figure~\ref{ex_bubbles} running on a bi-dual-core machine, this
leads to Figure~\ref{bulles_machine}: the bubble hierarchy is kept at
the top of the hierarchy of runqueues. Such a distribution permits
the use of all processors of the machine, since all of them
have the opportunity to run any thread.
This does not however take affinities between threads and processors
into account, since no relation between them is used.  On the
contrary, Figure~\ref{bulles_runqueues} shows how threads can be
spatially distributed in a very affinity-aware way.
However, if some threads sleep, the corresponding processors
become idle, resulting to a partial CPU usage.

We designed a programming interface~\cite{BubbleSchedAPI} for manipulating
bubbles and threads among the runqueues so as to achieve
compromises between distributions of Figures~\ref{bulles_machine}
and \ref{bulles_runqueues}.
Threads and bubbles are equally considered as \textbf{entities}, while
bubbles and runqueues are equally considered as \textbf{scheduling
  holders}, so that we end up with entities (threads or bubbles) that
we can schedule on holders (bubbles or runqueues). Primitives are then
provided for manipulating entities in holders.
Runqueues can be accessed through vectors, and can be walked thanks to
``father'' and ``son'' pointers.  Some functions permit to gather
statistics about bubbles so as to take appropriate decisions. This
includes the total number of threads and the number of
running threads, but also 
the accumulated expected and current CPU computation time or
memory usage, or the cache miss rates.  To handle concurrency, one can
use some fine-grain locking functions, but since bubble scheduling
decisions are generally rather medium-term oriented, one can also just
use a coarse-grain function which locks a whole sub-part of the machine
(runqueues and bubbles).  Once locked, one can enumerate entities held
by bubbles and runqueues, and redistribute them at will.  This permits
to elaborate complex manipulations without having to care about locking.

Writing a scheduler actually reduces to writing some hook
functions. The \texttt{bubble\_\-schedule()} hook is called when the
ground Self-Scheduler encounters a bubble while searching for the
next thread to execute. By default, it just finds a
thread in the bubble (or its sub-bubbles) and switches to it.
The \verb+bubble_tick()+ hook is called when a time-slice for a
bubble expires, permiting periodic operations on bubbles.
Of course, mere ``daemon'' threads can also be
started to perform background operations. 
Programmers may hence manipulate threads with a high level of abstraction by
deciding the placement of bubbles on runqueues.

\subsection{Implementation Examples of Scheduling Algorithms}

Our programming interface, though quite simple, permits to develop a
wide range of powerful ``bubble schedulers'', a few examples are
provided below.

\subsubsection{Burst Scheduling}

A first example is our previously-described
``burst algorithm''~\cite{THIBAULT:2005:31780}: a Self-Scheduling
algorithm ``pulls'' towards processors the bubbles, which ``burst''
(\emph{i.e.} release their content on a runqueue) in an opportunistic
way when they reach hierarchy levels given by programmers. To
implement this, the \texttt{bubble\_\-schedule()} hook
pulls the bubble a bit towards
the current processor, and its content is released if needed. In a few
iterations, bubbles and threads get distributed as on
Figure~\ref{bulles_runqueues}. Moreover, bubbles are periodically
``regenerated'' through the \texttt{bubble\_\-tick()} hook:
their original content is put back into them, and they are put back on
the machine runqueue for a new distribution. This permits to
dynamically adapt the distribution to new workloads while keeping
affinities in mind.  This quite simple algorithm was tested with
\emph{heat conduction and advection} simulations.  The results showed
that this permits to get the same performance benefit as manual thread
binding, but in a portable way.

\subsubsection{Gang Scheduling}
\label{gangsec}

In the 1980's, \textsc{Ousterhout} 
proposed to group data and threads
into \emph{gangs}, and schedule gangs on machines
instead of threads on processors. To realize such \emph{gang scheduler},
we express gangs thanks to bubbles, and a ``daemon thread''
performs the scheduling, see the dozen lines of code on
Figure~\ref{gang_code}. The daemon uses its own \verb+nosched_rq+ where it
puts all bubbles (i.e. gangs) aside, but leaves one on the
main runqueue for some time during which the basic Self-Scheduler
can schedule the threads of the bubble. The result is as expected:
time slices are equally distributed between gangs, and then threads
within gangs share their execution time within these time slices.
Programmers can easily tinker with this (change the
periodicity, etc.) without technical knowledge.

\begin{figure}[htb]
\centering
\begin{minipage}{7cm}
\scriptsize\tt
runqueue\_t nosched\_rq;\\
while(1) \{\\
\null\quad /* Wait for next time slice */ \\
\null\quad delay(timeslice);\\
\null\quad holder\_lock(\&main\_rq);\\
\null\quad holder\_lock(\&nosched\_rq);\\
\null\quad /* Put all entities aside */\\
\null\quad runqueue\_for\_each\_entry(\&main\_rq, \&e) \{\\
\null\quad\quad get\_entity(e);\\
\null\quad\quad put\_entity(e, \&nosched\_rq);\\
\null\quad \}\\
\null\quad /* Put one entity on main runqueue */\\
\null\quad if (!runqueue\_empty(\&nosched\_rq)) \{\\
\null\quad\quad e = runqueue\_entry(\&nosched\_rq);\\
\null\quad\quad get\_entity(e);\\
\null\quad\quad put\_entity(e, \&main\_rq);\\
\null\quad \}\\
\null\quad holder\_unlock(\&main\_rq);\\
\null\quad holder\_unlock(\&nosched\_rq);\\
\}
\end{minipage}
\hspace{0.1cm}
\begin{minipage}{4cm}
\scriptsize\tt
\begin{tabular}{rrrrr}
       name&pr &cpu\%&s &cpu \\
 gang sched&42 & 0.0 &I&0\\
        0-0&43 &26.4&R &3 \\
        0-1&43 &26.6&R &0 \\
        0-2&43 &26.8&R &2 \\
        0-3&43 &25.6&R &1 \\
        0-4&43 &26.6&R &3 \\
        1-0&43 &21.9&R &2 \\
        1-1&43 &22.2&R &0 \\
        1-2&43 &22.5&R &3 \\
        1-3&43 &22.1&R &2 \\
        1-4&43 &21.9&R &0 \\
        1-5&43 &22.6&R &1 \\
        2-0&43 &19.2&R &3 \\
        2-1&43 &18.9&R &1 \\
        2-2&43 &19.3&R &0 \\
        2-3&43 &19.8&R &2 \\
        2-4&43 &19.1&R &3 \\
        2-5&43 &19.3&R &1 \\
        2-6&43 &19.2&R &2 \\
\end{tabular}
\end{minipage}
\caption{A gang scheduler: source code and statistics
obtained with our \texttt{top}-like tool.}
\label{gang_code}
\label{gang_top}
\end{figure}

An interesting use of this gang scheduler is to emulate a network
of virtual machines on a single one in a fair way: each
virtual machine is represented by a bubble which contains its
thread. By running the gang scheduler, we get very fair
execution: each machine gets time slices
during which its threads can run
on the machine. The result are shown on Figure~\ref{gang_top}, where
three busy-looping gangs are sharing a four-processor machine (gang 0
has $5$ threads \verb+[0-4]+, etc.)

\subsubsection{Work-Stealing Scheduling}

One of our currently in-progress algorithms is based on work stealing:
the hierarchy of bubbles is first settled on the list of processor
0. Then, when \texttt{bubble\_\-schedule()} is called on an idle
processor, it will use our helper functions to look for work to steal
locally (on the runqueue of the other processor of the same chip for
instance), then more globally, until finding work to steal. The actual
work ``steal'' is a non-trivial algorithmic problem: only a part of
the bubble hierarchy should be pulled toward the idle processor,
and the structure of the hierarchy should be taken into account as
much as possible
All attributes and statistics attached to
bubbles should be carefully taken into account in heuristics, so as to
get a distribution suited to the application.
These are only purely algorithmic issues though: no technical problems remain.

\section{Implementation}
\label{sec:impl}

Our BubbleSched platform is currently implemented as an extension of
the \textsc{Marcel}
portable two-level thread library with a low additional overhead ($\sim
5\%$ on context switches~\cite{THIBAULT:2005:31780}).  This library
uses operating system functions to detect the machine architecture,
bind one kernel-level thread to each processor and it then
performs fast user-level context switches between user-level
threads. Therefore, assuming no other application is running, it
keeps complete control over thread scheduling on processors in
userspace.
Operations on bubbles can hence be also done in user space, and ``Daemon
threads'' are just \textsc{Marcel} threads.

Our BubbleSched platform also includes a debugger to help
understanding the behavior of a scheduler. A lightweight trace of
events (thread birth, sleep, wake up, bubble placement) is
recorded during the execution of the
application~\cite{DanNamWac05Europar} and analyzed
off-line. This trace can be converted into an animated movie that
shows interactively the series of scheduling events and placement
decisions that occurred during the execution. Thus,
programmers can easily replay the scheduling decisions at will to find
out their algorithmic flaws.

\section{Evaluation}
\label{sec:eval}

We show the usefulness of a programmable scheduling platform
by introducing an example that illustrates well the dilemma between
distributing threads over the machine or keeping related threads close
together. It was run on a bi-dual-core Opteron
system whose NUMA factor between the dual-cores is
around 1.4.

\begin{figure}[htb]
\centering
\includegraphics[scale=0.47]{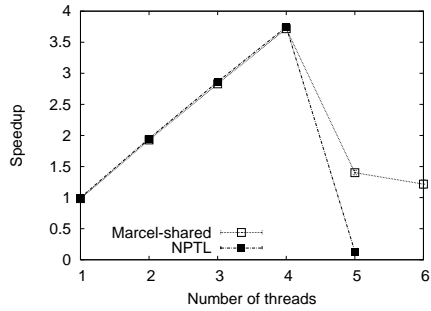}
\caption{\label{scaling}Parallel speedup of a single job}
\end{figure}

We experimented with SuperLU\_MT, a
thread-parallel solver for large, sparse systems of
linear equations (LU factorization). Figure~\ref{scaling} first shows
how well SuperLU scales on the target machine. The speedup is quite
close to perfect, up to the number of processors, but because of cache and synchronization affinities,
using a number of threads greater than the number of processors (4)
just makes performances really bad when using generic
schedulers like NPTL (\textsc{linux 2.6.17}) or original Marcel
with a single shared runqueue for all processors.

We consider a situation where a multi-scale scientific application
needs to perform LU factorizations jobs ``on demand''. There are many
ways to process these jobs, depending on how many threads to run per
job and how to schedule them.  Using a mere batch scheduler (running
each job 4-way up to completion), or a completely distributed
scheduler (running each job 1-way up to completion) may not be wise,
in case other parallel parts of the application (running on other
machines) need the job result sharply.

\begin{figure}[hbt]
\subfigure[4-way jobs] {
\includegraphics[scale=0.47]{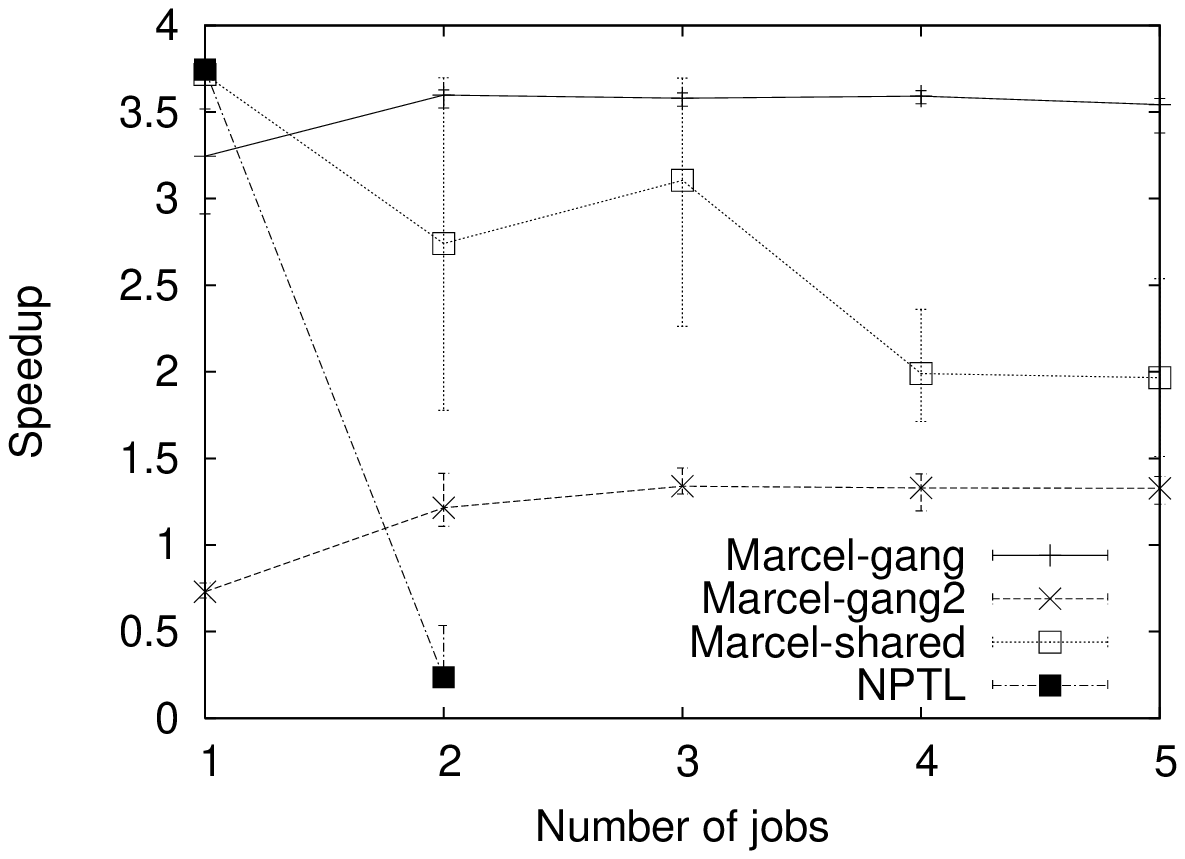}
\label{4way}
}
\subfigure[2-way jobs] {
\includegraphics[scale=0.47]{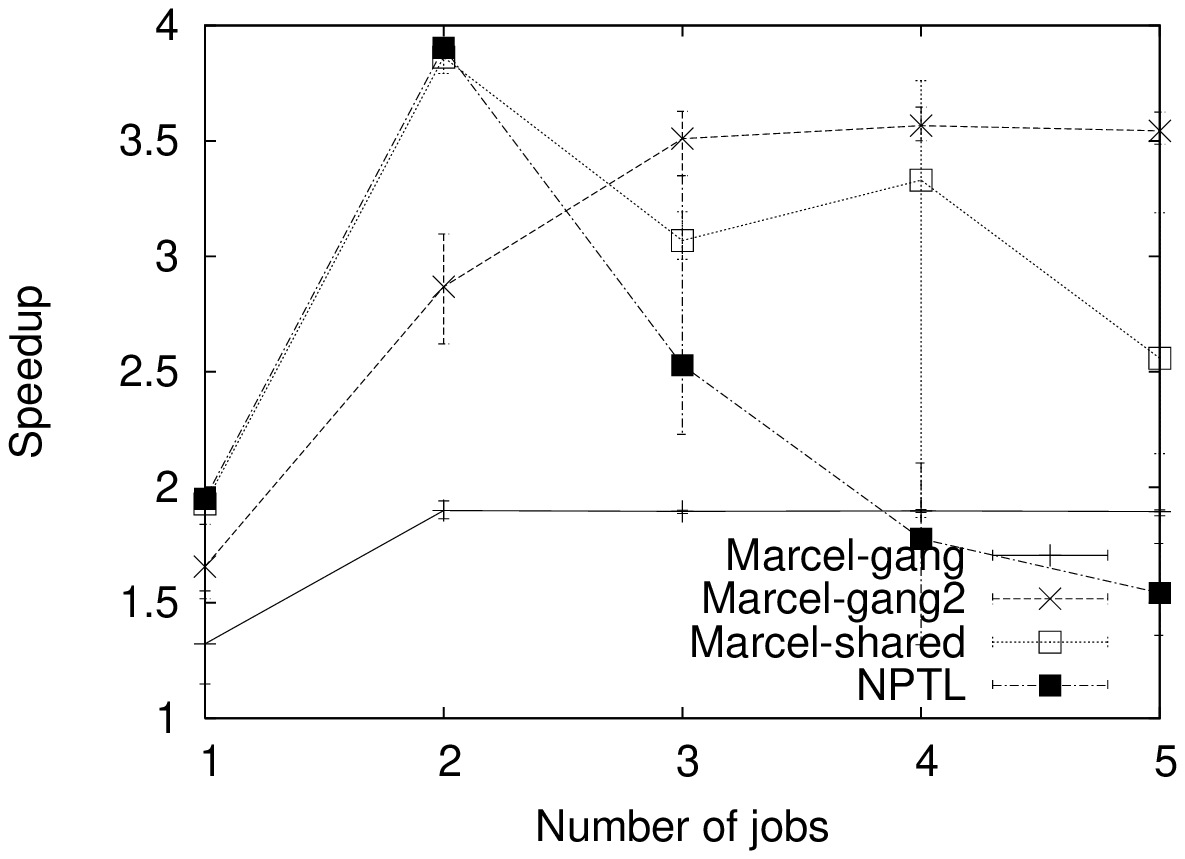}
\label{2way}
}
\caption{Parallel speedup of mixed jobs, using generic or gang scheduling policies}
\end{figure}

Figures~\ref{4way} and \ref{2way} show the results obtained using
several approaches. On figure~\ref{4way}, all jobs were run by using
four threads, while on figure~\ref{2way}, all jobs were performed
by using two threads.  Both figures clearly show that the generic
schedulers of NPTL and original Marcel get very bad and
erratic performances (min-max bars are very tall), because they equally
process all threads of all jobs without taking affinities into account. 
We used our gang scheduler of section~\ref{gangsec} for
running the jobs with a time slice of 0.2ms (the solving time of one
job with only one thread is around 10s) and a single
gang scheduler over the whole 4-way machine (``Marcel-gang'' curve),
but we also tried executing two gang schedulers, one
on each 2-way NUMA node of the machine (``Marcel-gang2'' curve).

The bottom curve of Figure~\ref{2way} shows that running only one gang
scheduler for 2-way jobs obviously limits the speedup to a value of
2. The bottom curve of Figure~\ref{4way} shows that running 4-way jobs
on 2-way NUMA nodes also get a limited speedup (because there are more
threads than processors). The top curve of Figure~\ref{4way} shows
that our gang scheduler performs quite well at running jobs on the
machine: each job seems to achieve a speedup of approximately 3.5,
whatever how many there are.  The ``Marcel-gang2'' curve of
Figure~\ref{2way} shows that the two gang schedulers achieve this
quite well too, provided that there are several jobs, of course.
Finally, by carefully comparing these last two curves, one can notice
that for this application on this machine, running one gang scheduler
for the whole machine is actually a little better than running two
separate gang schedulers on each NUMA node. This is probably because
we chose, for repartition reasons, to have these two gangs take and
put back jobs to a common job pool. Had the application been a little
different (less cache- but more memory-bound), we would have noticed the
converse.

It must be noted that this experiment was conducted without modifying
the application at all. We just made Marcel build bubbles according to
natural thread creation affiliation: since threads that work on the same
job are created by the same thread, the resulting bubble hierarchy
naturally maps to the job. It was then just a matter of starting gang
schedulers with appropriate parameters.

\section{Related Work}
\label{sec:rel-work}

Bossa~\cite{bossa} provides scheduling abstractions and a
language for developing schedulers.
However, its
goal is to \emph{prove} the correctness of the
scheduler, and as a such the proposed language, though powerful
enough for implementing the \textsc{Linux}~2.2 mono-processor
scheduler, is quite restrictive and limits programmers a lot.

\textsc{ELiTE}~\cite{LocalUserSched} provides an
a user-scheduling platform taking into account
affinities between threads, cache and data.
However, it lacks interaction with the application: affinities are
detected rather than provided by programmers.

Operating Systems have fairly good schedulers, and
Fedorova~\cite{fedorova} worked on cache-aware schedulers for
Operating Systems.  However, these are targetted towards ``blind''
multi-application situations and hence can't benefit from knowledge
provided by the programmer of a scientific application.

Several operating systems provide facilities for distributing kernel
threads along the machine by grouping them into sets: liblgroup on
Solaris, NSG on Tru64 and libnuma on Linux.
These look very much like single level bubbles, but no possibility of
nested sets is provided, which limits the affinity expressiveness.
Moreover, none of them provides the degree of
control that we provide: with \emph{BubbleSched}, the application has
hooks at the very heart of the scheduler to react to events like
\emph{thread wake up} or \emph{processor idleness}.

\section{Conclusion}
\label{sec:concl}

In this paper, we present the BubbleSched platform, a tool for
designing and prototyping specialized schedulers for specific
application domains and libraries (e.g. adaptive mesh refinement
algorithms, SPMD codes) for which threads behavior and memory
affinities can be predicted to some extent. It provides programmers
both with a way to express the application structure, and several
high-level scheduling distribution primitives that let scheduling
experts write \emph{bubble algorithms} that tightly ``drive'' the
thread scheduler by implementing some hooks.  Examples of implementing
scheduling strategies have shown how easy this is and how powerful it
can be. Non-expert programmers may try different combinations of
existing strategies to schedule threads, focusing on algorithmic
issues rather than on gory details. Actually, part of this work was
done in collaboration with researchers at the CEA (french Atomic
Energy Commission) who have been developing huge HPC applications for
a few decades and who are looking for a tool allowing them to transfer
their expertise to the underlying runtime system. The
conduction-advection application described in the paper is an example
of such application where we have demonstrated the interest of our
approach.

This work opens lots of future prospects. In the short term
several algorithmic approaches will be tested and tuned to schedule
real applications. On a longer run, a generic tunable scheduler
could take into account as much information as
possible from the hardware, the compiler and programmers. An
integration to the \textsc{Linux} kernel could even be considered,
since a bubble hierarchy naturally exists through the
notions of threads, processes, sessions and users.

\bibliographystyle{abbrv}
\bibliography{Bib/outils}
\end{document}